# Environmental Effects on Local Active Galactic Nuclei

Pierluigi Monaco[1,2], Giuliano Giuricin[1,2],
Fabio Mardirossian[1,2], Marino Mezzetti[1,2]

(1) Scuola Internazionale Superiore di Studi Avanzati (SISSA), via Beirut 4, 34013 – Trieste, Italy
(2) Dipartimento di Astronomia, Università degli studi di Trieste, via Tiepolo 11, 34131 – Trieste, Italy

email:
monaco@tsmi19.sissa.it
giuricin@tsmi19.sissa.it
mardirossian@tsmi19.sissa.it
mezzetti@tsmi19.sissa.it

## Abstract

Using an extensive sample of nearby galaxies (the Nearby Galaxies Catalog, by Tully), we investigate the environment of the galaxies hosting low-luminosity AGNs (Seyferts and LINERs). We define the local galaxy density, adopting a new correction for the incompleteness of the galaxy sample at large distances. We consider both a complete sample of bright and nearby AGNs, identified from the nuclear spectra obtained in available wide optical spectroscopic surveys, and a complete sample of nearby Seyferts. Basically, we compare the local galaxy density distributions of the AGNs with those of non-AGN samples, chosen in order to match the magnitude and morphological type distributions of the AGN samples.

We find, only for the early-type spirals more luminous than $\sim M^*$, that both LINERs and Seyferts tend to reside in denser environments on all the scales tested, from tenths of Mpc to a few Mpc; moreover Seyferts show an enhanced small-scale density segregation with respect to LINERs. This gives support to the idea that AGNs can be stimulated by interactions. On larger scales, tens of Mpc, we find that the AGNs hosted in luminous early-type spirals show a tendency to stay near the center of the Local Supercluster. Finally we discuss the interpretations of our findings and their consequences for some possible scenarios of AGN formation and evolution and for the problem of how AGNs trace the large-scale structures.

*Subject headings:* galaxies: active — galaxies: clustering — galaxies: interactions — galaxies: nuclei — galaxies: Seyferts

# 1 Introduction

It was suggested long ago that the galaxy environment can have an important role in feeding nuclear galactic activity; this issue has received ever growing attention in the last fifteen years. Here, by nuclear activity, we mean the presence of nuclear emission lines that cannot be explained in terms of normal stellar populations; moreover, we will concentrate on low-luminosity (Seyfert and LINER) activity, i.e. the kind of nuclear activity that can be found in the volume of the nearby universe studied in this paper.

Interactions, which are generally revealed by the presence of nearby companions or the appearance of morphological distortions, were considered as being possibly responsible for Seyfert nuclei (e.g. the first observational studies of Vorontsov-Velyaminov 1977; Adams 1977; Simkin, Su & Schwartz 1980) and even for more energetic objects, such as radio galaxies and QSOs (see the review by Balick & Heckman 1982). To assess observationally the role of interactions on AGN activity, it was necessary to verify if AGNs tend to have nearby companions or distorted morphologies more frequently than normal galaxies have, or, equivalently, to verify if samples of paired or distorted galaxies show a greater fraction of AGNs with respect to non-interacting "field" galaxies. Efforts in this direction are reviewed by Fricke & Kollatschny (1989), Heckman (1990) and Osterbrock (1993).

The first extensive study of the environment of Seyfert galaxies was carried out by Dahari (1984). Using the Palomar Observatory Sky Survey, he estimated the background-corrected number of neighbours (within 3 projected diameters) of a sample of 103 Seyferts and of a control sample of normal galaxies. The latter was constructed in the following way: for every Seyfert he chose three galaxies within 3° and with apparent diameters between 0.75 and 1.5 times the diameter of the Seyfert. The percentage of Seyferts with physical companions turned out to be at least 5 times that of the comparison sample. Similar results were reported by MacKenty (1989) for Markarian Seyferts, with respect to a comparison sample chosen in a way similar to Dahari's (1984). On the other hand, Markarian Seyferts showed no difference in environment with respect to Markarian non-Seyferts, which are mostly starburst galaxies. Using an approach similar to Dahari's (1984) and more recent data, Rafanelli & Violato (1993) found again an excess of physical companions for Seyferts. Finally, many authors have suggested that Seyfert 2 galaxies alone could be responsible for most of the claimed environmental effect (Petrosian 1982; Dahari & De Robertis 1988; MacKenty 1989; Simkin 1990). Yet two factors at least may have contributed to these results: the Seyfert sample is biased toward early-type spirals, which are known to reside preferentially in denser environments than late-type ones do (see, e.g., Giuricin et al. 1988; Tully 1988b). Secondly, Seyferts tend to be hosted in luminous galaxies (see, e.g., Danese et al. 1992), so that a comparison sample, constructed to have roughly the same apparent diameter/magnitude distribution, is likely to be made of less luminous and thus nearer galaxies with smaller physical sizes; this could cause a greater number of companions to be erroneously assigned to Seyferts.

Besides, much effort has been devoted to check whether a larger fraction of Seyferts occurs in samples of paired or distorted galaxies. Positive results were reported by Dahari (1985a), who examined the frequency of Seyferts in a subsample of the Vorontsov-Velyaminov Catalogue of Interacting Galaxies (1959, 1977), although no Seyferts were found in the extremely interacting systems. Similarily, Kennicutt & Keel (1984) and Keel et al. (1985) noted an excess of Seyferts in samples of paired and Arp (1966) galaxies; moreover, Keel et al. found the nuclear emission lines of Seyferts in the paired and Arp samples to be enhanced with respect to those of the comparison sample, a correlation which was improved by adding LINERs to the Seyfert sample. Finally, Laurikanen & Moles (1989) found a high fraction of their sample of interacting galaxies to show LINER activity.

On the other hand, several studies have not confirmed the importance of the environment for AGNs. Fuentes-Williams & Stocke (1988) analysed the environment within 1 Mpc of a sample of 53 Seyferts and of a comparison sample of 30 galaxies, chosen in order to match the Hubble type, luminosity and redshift distributions of the Seyfert sample, and to have linear diameters in a given range (in order to avoid foreground-background contamination). They found no environmental difference between Seyferts and comparison galaxies. Dropping the lower limit in the linear size of the companions, they found a small but significant tendency of Seyferts to have more companions. This find-ing and Dahari's (1984) one can be reconciled if Dahari's positive result is mostly due to the biases discussed before and if the physical effect is given mostly by small companions, with $M_B \geq -18$ (see the discussion in Fuentes-Williams & Stocke 1988). Furthermore, at variance with Keel et al. (1985), Dahari (1985b) and Dahari & De Robertis (1988) found no relevant correlations between nuclear activity and interaction parameters. Bushouse (1986) noted that Seyferts and LINERs are rare ($\sim$1%) in violently interacting galaxies. More recently, Sekiguchi & Wolstencroft (1992, 1993) did not find any excess of Seyferts or LINERs in IRAS-selected strongly interacting objects.

Also the cluster environment seems to play a significant role in determining the nuclear activity of galaxies. Gisler (1978) noted that emission-line galaxies are not common in rich clusters; this was interpreted as the effect of ram-pressure sweeping of galactic gas. Petrosian (1982) found that Seyferts follow the general tendency of galaxies to cluster but they avoid rich clusters. Dressler, Thompson & Shectman (1985) confirmed that AGNs are rare ($\sim$1%) in rich clusters. Nesci (1986) found that the cluster environment of Seyferts is different from that of ellipticals but very similar to that of late-type spirals; he attempted to explain this by means of a morphological evolution, driven by AGN activity, from late to early spirals. Finally, Petrosian & Turatto (1986) showed that Markarian galaxies (often Seyferts) prefer medium-compact clusters, but again tend to avoid dense cluster cores.

Summing up, many authors have claimed a correlation between interactions and the incidence of Seyferts, but the question is still hotly debated; it seems that the environment can be effective but neither necessary nor sufficient in producing nuclear activity. Besides, the role of LINERs is particularly unclear, mainly because they are very difficult to observe in galaxies outside the Local Supercluster.

Also the theoretical arguments concerning the importance of interactions in stimulating nuclear activity have not been fully assessed. Non-axisymmetric perturbations stimulated by interactions, such as a bar, a trailing spiral density wave or the accretion of small satellites, can be efficient in making some gas lose angular momentum and in moving it to the galactic central kpc (Shlosman 1990; Shlosman, Frank & Begelman 1989; Shlosman, Begelman & Frank 1990; Lin, Pringle & Rees 1988; Barnes & Hernquist 1992). Many numerical simulations essentially confirm these ideas (Byrd, Smith & Miller 1984; Byrd, Sundelius & Valtonen 1987; Noguchi 1988a,b; Hernquist 1989a,b, 1990). Moreover, Byrd et al. (1986) found that their fraction of simulated interacting galaxies with a significant infall of gas into the nucleus roughly matched the fraction of Seyferts found in interacting systems by Dahari(1985b).

When the gas has fallen into the central kpc, the tidal force is no longer efficient in making it fall further inside. In the nuclear region the gas will presumably give rise to bursts of star formation, unless some other mechanisms — such as nuclear bars, magnetic torques, cloud-cloud collisions or supernovae driven turbulence — forces it to fall toward a central supermassive black hole (Lin, Pringle & Rees 1988; Shlosman et al. 1989, 1990; von Linden et al. 1993). Observational evidence of gas infall into Seyfert nuclei is reviewed by Heckman (1992).

The present, unclear state of knowledge about the influence of the local environment on the nuclear activity of galaxies induced us to examine the problem. In this paper we analyze the local density distribution of a complete sample of AGNs in the Local Supercluster. In § 2, using the Nearby Galaxies Catalog (Tully 1988a) to characterize the 3D galaxy distribution within the Local Supercluster, we define a set of local galaxy density parameters, paying attention to the correction for catalog incompleteness, which is different from that adopted in previous relevant works (see, Tully 1988b; Giuricin et al. 1993a, 1994). In § 3 we present the samples of AGNs used in this paper, and in § 4 we closely analyze the distributions of local densities of AGNs and non-AGNs at various scales (and of other related parameters), finding a significant segregation of AGNs both on small scales (tenths of Mpc) and on large scales (tens of Mpc). Finally, in § 5 we summarize our results and briefly discuss their implications for some possible scenarios of AGN formation and evolution.

## 2 The Local Galaxy Density

From the discussion outlined above, it is clear that the major problems in dealing with environmental effects on galaxy properties are the definition of the samples to be compared (AGNs and non-AGNs in our case) and a clear characterization of the environment. In agreement with Heckman (1990), we prefer to look for nearby companions rather than for distorted morphologies as a sign of possible interactions; in fact there is always the possibility that some distortion is caused by the nuclear activity itself, or in any case by internal dynamics.

A rigorous assessment of the environment is often lacking: only the projected distances are known so that statistical background corrections are needed; the maximum radius at which a galaxy is considered to be a companion is arbitrary; moreover, taking into account only the percentage of objects with companions in a sample means having a single measure for every sample, which is statistically much less advantageous than assigning every galaxy an environmental parameter.

As already done in recent studies regarding the environmental effects on the bars (Giuricin et al. 1993a) and Arm Classes (Giuricin et al. 1994) of nearby spirals, we use the Nearby Galaxies Catalog (Tully 1988a; NBG hereafter) to give a 3D definition of environment. This catalog is intended to include all the known nearby galaxies with systemic velocities of less than 3000 km/s, which corresponds to a distance of 40 Mpc with the Hubble constant $H_0$=75 km s$^{-1}$ Mpc$^{-1}$, the value adopted throughout the present paper. In the NBG catalog every galaxy is given a distance, based on its redshift, on the assumed value for $H_0$ as given before, and on corrections for Virgo infall and group membership. The correction for Virgo infall is made by means of the Virgocentric retardation model described by Tully & Shaya (1984), in which the authors assume the Milky Way to be retarded by 300 km/s from the universal Hubble flow by the mass of the Virgo Cluster. The correction for group and cluster membership is made by assigning every group member a distance consistent with the mean velocity of the group itself (see NBG and references therein).

With this 3D distribution of galaxies we can define a parameter of local galaxy density: following Tully (1988b) we define the parameter $\rho_\sigma$ (in galaxies per Mpc$^3$) as the number of galaxies per Mpc$^3$ that are found around every galaxy within the smoothing length $\sigma$ (in Mpc):

$$\rho_\sigma = \sum_i C \exp[-r_i^2/2\sigma^2], \quad (1)$$

where every galaxy is smoothed with a gaussian filter of half-width $\sigma$, $r_i$ is the spatial distance of the $i$-th galaxy from the specified galaxy and the normalization coefficient is $C = 1/(2\pi\sigma^2)^{3/2} = 0.0635/\sigma^3$;

the sum is carried over all galaxies except the one we are calculating the density for.

The definition given above does not take into account the increasing incompleteness of the catalog at large distances. Tully (1988b) found a smooth curve $F(\mu)$ – where $\mu = 5\log(R) + 25$ is the distance modulus and the distance $R$ is in Mpc – to express the number of galaxies brighter than $M_B = -16$ that exist for every galaxy catalogued. To correct for incompleteness, he multiplied the smoothing length by a factor of $F^{1/3}$, so that the galaxies enclosed in the enlarged volume could account for the ones missing in the catalog. This kind of correction neglects the fact that galaxies tend to cluster: in Giuricin et al. (1993a) we showed that the mean value of the density $\rho_\sigma$ scales with $\sigma$ as

$$\langle \rho_\sigma \rangle \propto \sigma^{-\gamma}, \quad (2)$$

where a power law for the two-point correlation function is assumed, $\xi(r) \propto r^{-\gamma}$ (to be precise, Eq. (2) is valid in the limit $\langle \rho_\sigma \rangle \gg N_{tot}/V$). As a result, if we correct for incompleteness by increasing $\sigma$ by $F^{1/3}$ we get densities systematically lower, on average, by a factor $F^{-\gamma/3}$, which is 0.74 at 20 Mpc, 0.64 at 30 Mpc and 0.39 at 40 Mpc (for $\gamma \sim 1$ in our sample; see Giuricin et al. 1993a). Then we decided to correct the density simply by weighting every galaxy by $F$; this increases the statistical errors due to the decreased number of objects, but does not introduce any bias. Moreover, we prefer to use a new expression for $F(\mu)$, which fits somewhat better the observed incompleteness at increasing distances:

$$F = \exp[0.033(\mu - 28.5)^{2.83}] \quad (3)$$

(and $F = 1$ when $\mu < 28.5$). Finally, our density parameter, corrected for incompleteness, is:

$$\rho_\sigma = \sum_i C F(\mu_i) \exp[-r_i^2/2\sigma^2]; \quad (4)$$

$\rho_\sigma$ gives the number of galaxies, per Mpc$^3$, brighter than $M_B = -16$ around the galaxy considered.

As discussed in Giuricin et al. (1993a), because of the clustering properties of galaxies, the choice of different $\sigma$-values implies a different physical meaning for the local galaxy density $\rho_\sigma$, so that to understand the behaviour of a sample we need to examine its $\rho_\sigma$ distributions for different values of $\sigma$. The density distributions of different galaxy samples will be compared by means of the two-tailed non-parametric Kolmogorov-Smirnov (KS), Rank-Sum (RS) and Mann-Whitney (MW) tests, which give the probability that two distributions are not random realizations of the same parent distributions (see, e.g., Hoel 1971). The KS test statistics is based on the maximum difference between two cumulative distributions, while the RS and MW tests are both based on the ordering of measures, and are more efficient in discovering systematic differences between two distributions.

In order to calculate the local density $\rho_\sigma$ we need to know the absolute magnitudes of the NBG galaxies, as the ones with $M_B \leq -16$ will be taken to be contributors to $\rho_\sigma$. For the NBG galaxies which do not have $M_B$ tabulated in the catalog (according to the distances adopted and the corrected total blue apparent magnitudes) we have estimated it from their corrected isophotal diameters $D_{25}$ (relative to the 25 B mag arcsec$^{-2}$ brightness level), by relying on the following standard luminosity-diameter relations:

$$M_B = -4.8 \log D_{25} - 13.8 \quad (5)$$

with $D_{25}$ expressed in kpc, for elliptical galaxies (Giuricin et al. 1989) and

$$M_B = -5.7 \log D_{25} - 12.4 \quad (6)$$

for lenticular and spiral galaxies (Girardi et al. 1991).

Finally, we have updated the Hubble morphological types and the bar-types (SA=unbarred, SB=barred, SAB=transition-type) of the NBG galaxies by consulting the Third Reference Catalogue of Bright Galaxies (de Vaucouleurs et al. 1992).

## 3 The Samples of AGNs

In order to define in a suitable way a sample of AGNs and a comparison one of non-AGNs, we have joined together three spectroscopic optical surveys of the nuclei of complete samples of bright galaxies. The first one is the optical (and radio) survey of Heckman, Balick & Crane (1980; HBC80 hereafter). They observed the nuclei of 88 galaxies, taken from the Second Reference Catalogue of Bright Galaxies (de Vaucouleurs, de Vaucouleurs & Corwin 1976), and selected according to $B_T \leq 12$ and $\delta \geq 40°$. The optical observations were performed with an aperture diameter of 6 arcsec, a resolution of 8 Å and an integration time of 12 to 20 min. They tabulated the fluxes of the principal emission lines, like H$\alpha$, H$\beta$, [OII]$\lambda$3727, [OIII]$\lambda$5007, [OI]$\lambda$6300, [SII]$\lambda\lambda$6717,6731 and [NII]$\lambda$6584, upon which Heckman (1980a,b) based his classification of types of nuclei.

The second survey, by Keel (1983, hereafter K83), was performed to complement the HBC80 survey in the northern sky: all galaxies (with morphological types between S0/a and Scd) were observed with $B_T \leq 12$ and in the area of the sky with $-15° \leq \delta \leq 40°$. The spectroscopical observations were performed with apertures ranging from 8.1 to 4.7 arcsec; the lines used for the classification were basically H$\alpha$, [NII]$\lambda$6584 and [SII]$\lambda\lambda$6717,6731. The observation time was set long enough to detect emission lines; in this way the percentage of emission nuclei is higher in K83 than in other two surveys, which were carried out with fixed exposure times.

Thirdly, Véron-Cetty & Véron (1986; VV86 hereafter) studied the spectra of the nuclei of a complete southern sky sample of galaxies. They selected all the galaxies from the Revised Shapley-Ames Catalogue (Sandage & Tammann 1981) with $\delta \leq 20°$, $cz \leq 3000$ km/s and $M_B \leq -20.1$ (with our $H_0$); apart from the sky region observed, the selection criteria are only slightly different from those of HBC80 and K83. The telescope aperture was 4x4 arcsec, the resolution 10 Å and the exposure time 20 min. The emission lines used for the classification were H$\alpha$, H$\beta$, [NII]$\lambda$6584, [OI]$\lambda$6300, [OIII]$\lambda$5007 and [SII]$\lambda\lambda$6717,6731.

After having checked the consistency of the classifications made by the three authors (who substantially rely on the classification precepts of Baldwin, Phillips & Terlevich 1981, which are based on the relative strengths of suitable emission lines), we combined the three samples into a single one, which we shall call the All Sky Sample (ALSK). In order for the sample to be properly selected, we took from VV86 only those galaxies with $B_T \leq 12$, and from HBC80 and K83 only galaxies with $M_B \leq -20.1$. Moreover, we decided to cut our volume at a distance of 36 Mpc, beyond which incompletion becomes too severe and border effects start to affect our densities significantly.

In the area $-15° \leq \delta \leq 20°$ the two samples VV86 and K83 overlap; we chose to give priority to VV86 because of the larger volume investigated, the smaller apertures used in the observa-tions (which ensure a smaller degree of contamination by circumnuclear radiation), and the fixed exposure time, which avoids the problem discussed before. Nonetheless, the galaxies in common in the two samples allow us to verify that the classifications of the two authors are in substantial agreement, except that 7 nuclei, out of the 53 in common, were classified as emission line-free by VV86 and as LINERs by K83. Of these, six are in our ALSK sample: four of them have, according to K83, an equivalent width (EW) of [NII]$\lambda$6584 smaller than 1.5Å, the stated detection limit of VV86, while the other two have a slightly higher EW. So, to make VV86 and K83 fully consistent, we have to consider carefully the K83 galaxies with an EW smaller than or about 1.5Å.

ALSK contains 259 galaxies, 212 of which are from VV86, just 9 from K83 (the zone of overlapping with VV86 contains the Virgo Cluster, and thus most of the K83 galaxies) and 38 from HBC80; of the 9 galaxies from K83, just one is listed with EW of [NII]$\lambda$6584 smaller than 1.5Å, so that the the greater sensitivity of emission-line detection of K83 ought not to pose any problem. Besides, the absence of early- and late-type galaxies in K83 is not a severe problem, owing to the small number of K83 galaxies (moreover, we shall find non-null results only for early-type spirals).

We divided our ALSK galaxies into two subsamples: 1) the AGN sample, which includes Seyferts 1 and 2, LINERs, galaxies classified as N by VV86, which are Seyferts 2 or LINERs, and objects with composite spectra, which show clear sign of AGN plus star-formation activity; 2) the comparison sample, which includes emission line-free nuclei, HII region-like nuclei and a few uncertain or unclassified cases, which do not present sure evidence of AGN activity. We have attempted to divide the AGNs into two subclasses, the few sure Seyferts and the other AGNs, which include both LINERs and other objects whose distinction between LINER and Seyfert 2 is not possible on the basis of the available spectra; as the latter subclass is likely to contain mostly LINERs, we shall denote as Seyferts and LINERs the two subclasses. Unfortunately, the small number of Seyferts will not allow us to reach strong statistical conclusions about the different behaviours of the two kinds of AGNs.

We stress that the definition of ALSK is not based on nuclear activity, but rather ALSK is a well-defined sample limited in magnitude, absolute luminosity and volume; this ought to minimize any selection effect. The nuclear classification has been made by different authors, but through essentially the same criteria; moreover, ALSK is dominated by the VV86 sample, so that small differences in the classification ought not to give severe problems. Nonetheless, we shall check below, at the end of §4.1, the robustness of our results with respect to the nuclear classification and to selection effects due to luminosity, distance and galactic declination.

In order to construct a wider sample of Seyferts, we used the fifth edition of the Catalogue of Quasars and Active Nuclei (Véron-Cetty & Véron 1991), which contains a compilation of all the galaxies known as Seyferts in the literature. As far as bright Seyferts are concerned, this catalogue is thought to be essentially complete up to $z < 0.1$ (Rafanelli & Violato 1993). Moreover, it contains a few LINERs, but without any guarantee of completeness. From this catalog we have extracted all the galaxies in common with NBG, within 36 Mpc and catalogued as S1 or S2; we will call VV91 this sample of 46 Seyferts.

NBG allows us to define a suitable comparison sample for VV91. We want our comparison sample to have the same distribution in absolute and apparent magnitude and morphological type as the Seyfert sample. Therefore, we have used all the 484 NBG galaxies (not belonging to VV91) with $M_B \leq -19$, $B_T \leq 12$ and, of course, within 36 Mpc. In this way the $M_B$ distribution of VV91 differs by less than 80% from the comparison one. The $B_T$ distributions actually differ slightly but significantly; the same happens when we take all NBG galaxies with $B_T \leq 12.5$. However, we have noted that the $\rho_\sigma$ distributions of the comparison sample – as well as the distributions of the distances from the Virgo cluster, which will be used below – are quite stable with respect to the choice of the $B_T$ cutoff: using a cutoff of 12 or 12.5 does not affect significantly (even at the $1\sigma$ level) our $\rho_\sigma$ distributions, despite the very good statistics. We conclude that the slight mismatch in the Seyfert and comparison $B_T$ distributions does not significantly affect the $\rho_\sigma$ distributions, and thus it is not a serious problem for the kind of analysis we have carried out. Finally, we will always divide our sample into morphological type subsamples, in order to investigate separately the quite different behaviours of the different types, and also to avoid any bias due to the different type distributions which characterize the Seyferts and the comparison galaxies.

# 4 Analysis and Results

## 4.1 The All Sky Sample

It is well known that the presence of an AGN in a galaxy is a function of the galaxy morphology, and that the morphological type of a galaxy correlates strongly with the local environment, early types being located in denser regions. As a consequence, a relation between local environment and nuclear activity can be found as a spurious result of the correlations mentioned above. To avoid this bias we have divided ALSK into three different type subsamples: ellipticals and lenticulars (E+S0), early-type spirals (from S0/a to Sb; hereafter $S_{early}$) and late-type spirals (later than Sb) plus irregulars ($S_{late}$). We stress that this division is not made *a priori*, but in order to isolate the subset of types responsible for all the results found below. In Table 1 we list the number of objects in every subsample.

For the three subsets just defined we have compared the $\rho_\sigma$ distributions of AGNs and non-AGNs, for $\sigma = 0.25$, 0.5, 1.0 and 2.0 Mpc. No difference has been found for the E+S0 and $S_{late}$ galaxies. Although AGNs tend to reside in $S_{early}$ galaxies, the lack of any density segregation in the other type subsamples is not likely to be caused simply by poor statistics. As the independence of AGN activity in E+S0 and $S_{late}$ with respect to the environment has been confirmed by all the analyses we have carried out, in the following we will concentrate mainly on the early-type spirals.

In Fig. 1 we present the $\rho_\sigma$ distributions of $S_{early}$ AGN and non-AGN galaxies (Seyferts and LINERs are also shown separately), and in Table 2 we report the significance of the differences between the two distributions. It is evident that AGNs tend to reside in denser environments on substantially all the scales tested, from 0.25 Mpc, a scale at which we expect to see the effects of interactions with close companions, to 2.0 Mpc, a scale at which the cluster environment becomes important. The significance of the segregation is generally high, except for the case $\sigma = 0.25$ Mpc, where it is still $> 2\sigma$ (according to the RS and MW tests).

The small scale ($< 1$ Mpc) segregation of AGNs provides us with evidence of the role of tidal interactions in stimulating nuclear activity; if they were the dominant effect, we would expect the significance of the segregation to be lower at larger scales, as happened for the density segregation between barred and unbarred early-type spirals (Giuricin et al. 1993a). Instead, we see a strong, significant segregation of AGNs at scales as large as 2 Mpc. This could be due to an excess of AGNs in the Virgo cluster, where $\rho_\sigma$'s are the largest for $\sigma \geq 1.0$ Mpc. In fact, of the 9 $S_{early}$ Virgo galaxies (as assigned by NBG) in ALSK, 8 are classified as AGNs. Taking them out of the sample, all the segregations remain, though with a lower significance (see Table 2); so our findings are not simply a consequence of this excess of AGNs in the Virgo cluster.

As the $\sigma \geq 1.0$ Mpc segregation is not simply due to the observed excess of AGNs in Virgo, it could be caused by a different preferred location of AGNs within the Local Supercluster. We know (e.g. Tully 1982) that our Local Supercluster is characterized by a rough cylindrical symmetry around the Virgo cluster and the supergalactic plane. Thus, physically interesting quantities are the distance from the Virgo cluster, $D_V$, and the cylindrical virgocentric supergalactic coordinates, i.e. the distance from Virgo on the supergalactic plane $R_{SG}$, the polar angle $\Theta_{SG}$, and the distance from the plane $Z_{SG}$. Fig 2 shows the distributions of $D_V$, $R_{SG}$, $\Theta_{SG}$ and the modulus of $Z_{SG}$ – we have checked that the $Z_{SG}$ distributions are symmetrical with respect to the supergalactic plane – for AGN and non-AGN samples; Seyferts and LINERs are not shown separately as they always show essentially the same distribution. As regards the $D_V$ distributions, we can see from Fig. 2 and from Table 2 that AGNs tend to stay around the Virgo cluster on scales of tens of Mpc, a $\sim 3\sigma$ difference according to the RS and MW statistics; moreover, we note that the maximum difference is between 15 and 35 Mpc, and that it remains significant when the excess of Virgo AGNs is removed (see Table 2). From Fig. 2 and Table 3, we see that this difference is slightly more pronounced on the supergalactic plane ($R_{SG}$ distributions) than perpendicularly to it ($Z_{SG}$ distributions) and, in agreement with the symmetry of the Local Supercluster, there is no significant difference in the $\Theta_{SG}$ distributions.

As a consistency check, we have extracted from NBG all the $S_{early}$ galaxies with $M_B \leq -20.1$, $B_T \leq 12$, within 36 Mpc and not classified as AGNs in our ALSK sample. This new non-AGN sample consists of 48 galaxies; the comparison of its $\rho_\sigma$ and $D_V$ distributions with AGN ones confirms all the density segregations found above, but with improved statistical significance (see Table 2). However, we shall continue using our well-defined ALSK comparison sample, in order to minimize all the possible biases.

In Fig. 1 we also show the density distributions of the two subclasses of $S_{early}$ AGNs, the 11 Seyferts and 44 LINERs: both follow essentially the same distribution, according to the usual KS, RS and MW tests. Nonetheless, on smaller scales the Seyferts seem to prefer denser regions than LINERs do; some evidence for this can be found in the fact that on smaller scales the 11 Seyferts alone differ more significantly than LINERs alone from the non-AGN sample (see Table 2).

We wonder now whether the segregations found are characteristic of the inner zones of the Local Supercluster or not. We have divided our sample into galaxies with $D_V$ less and greater than 26 Mpc; we have found that this division has no effect on the segregations of AGNs in $\rho_\sigma$ and $D_V$ on all scales, except that small scale segregations are found to be stronger in the inner parts of the Supercluster simply because of the presence of the Virgo galaxies. In the same way we have tried to divide our objects into ones brighter and fainter than $M_B = -20.3$, obtaining similar results in the two cases. Unfortunately, the division into subsamples lowers the statistics, so that fine effects cannot be detected with such a small number of objects.

Instead, interesting results are obtained if we subdivide our objects according to the presence of a bar. We know from Giuricin et al. (1993a) that early-type barred spirals prefer denser environments on scales of 0.25 – 0.5 Mpc with respect to unbarred counterparts. If AGNs were hosted preferentially in barred spirals, the small-scale segregation found could be explained in these terms. Actually, we find that 18 of our $S_{early}$ AGNs are hosted in SA, 17 in SAB and 19 in SB systems, while the incidence of non-AGNs in the bar subsamples is 11, 8 and 9; as a consequence, our small-scale results cannot be caused by a density–bar relation. Nonetheless, when we subdivided our $S_{early}$ sample into non-barred (SA+SAB) and barred (SB) galaxies, we found no small-scale segregation for the barred ones, whereas it was enhanced for the unbarred objects alone (35 AGNs and 19 non AGNs; see Table 2). We conclude that a significant (>99%) small-scale density segregation of AGNs hosted in $S_{early}$ is present when a bar perturbation does not intervene to inhibit it.

Next, we investigated the robustness of our findings with respect to the nuclear classification. We know that essentially all galactic nuclei show some emission lines (Keel 1983), so that different detection limits may end up in different classifications. Being aware of this problem, we tried to consider as being AGNs only those whose EW of [NII]$\lambda$6584 is greater than 2.5Å; in this way we had 42 AGNs and 42 non-AGNs. As a result, the $\rho_\sigma$ and $D_V$ segregations are essentially confirmed on both small and large scales. Moreover, we have verified that changing the classification of a few uncertain cases does not affect any of our results.

Finally, we checked whether our results show a dependence on observer-dependent parameters, namely the apparent magnitude, the distance from us and the galactic declination $b$ (i.e., the angular distance from the avoidance zone of the Milky Way). As a matter of fact we find that, while the $M_B$ distributions are the same for $S_{early}$ AGNs and non-AGNs, the $B_T$ distributions are slightly (~94%) different, AGNs being hosted in brighter galaxies. This selection effect ought not to pose any problem for two reasons: first, we have seen before that the $\rho_\sigma$ distributions, as well as the $D_V$ ones, are insensitive to variations in the $B_T$ distributions; second, this selection effect becomes non-significant when we take as AGNs only those with EW ([NII]$\lambda$6584) > 2.5 Å (as done before), a case for which all our findings are confirmed. Next, we have verified the robustness of our findings with respect to errors in the correction for catalog incompleteness by checking the validity of our results for the $S_{early}$ galaxies within 30 Mpc, beyond which incompleteness becomes severe. Finally, we have verified that the percentage of $S_{early}$ AGNs and non-AGNs near the avoidance zone ($|b| < 20°$) is the same (~10%), so that no bias is to be expected from a selection effect connected to Milky Way obscuration.

Before going on, we wish to mention the spectroscopic survey of a complete sample of galaxies carried out by Huchra & Burg (1992), who classified the nuclei of the galaxies of the CfA Redshift Survey (Huchra et al. 1983). They identified a sample of Seyfert 1 and 2 galaxies complete to $B_T \leq 14.5$ and a sample of LINERs complete to $B_T \leq 12$. We have extracted from that list the 37 Seyferts and LINERs in common with NBG and within 36 Mpc;

their magnitude distributions show a cutoff roughly at $B_T = 12$ and $M_B = -19$, so we have used as a suitable comparison sample all the NBG galaxies in the same area of the sky and with the same limits of apparent and absolute magnitude. Only the subset of 12 $S_{early}$ AGNs with $M_B \leq -20.1$ shows some density segregation, not the whole sample of 18 $S_{early}$ AGNs. This is consistent with our previous findings, but not independent, as all the luminous and nearby CfA AGNs are contained in ALSK. At the same time, we get the further, independent suggestion that no density segregation is present for low luminosity $S_{early}$ spirals (with $M_B \geq -20$).

## 4.2 The VV91 Sample

Our VV91 sample contains a number of Seyfert 1 and 2 galaxies, but excludes all the LINERs. Before going on, we may wonder what we expect to observe on the basis of our previous findings. We have seen that our ALSK AGN sample is dominated by LINERs, about four times more numerous than Seyferts. So, when we take a sample of Seyfert galaxies alone, a suitable comparison sample will contain a large number of LINERs, and statistical differences between the comparison and Seyfert $\rho_\sigma$ and $D_V$ distributions will be very hard to detect, unless Seyferts segregate more than LINERs. So we expect our analysis of the VV91 sample to give us information about the previously suggested excess of small-scale segregation of Seyferts with respect to LINERs. On a large scale, Seyferts ought to behave like LINERs, so we expect to see no segregation.

Again we divided our sample into the same subsets of different morphological types as before, i.e. the E+S0 galaxies, the $S_{early}$ and the $S_{late}$ ones. As before, no segregation is observed, in $\rho_\sigma$ and in $D_V$, between Seyferts and non-Seyferts, for the E+S0 and $S_{late}$ subsets. For the $S_{early}$ (28 Seyferts and 128 non-Seyferts) we find no significant segregation in $\rho_\sigma$ until, as in the study of the CfA sample of AGNs, we restrict the sample to early spirals with $M_B \leq -20.1$ (incidentally the same limit as the ALSK sample), finding some small-scale segregation; nothing is observed on larger scales and on $D_V$. Fig. 3 shows their $\rho_\sigma$ distributions for $\sigma = 0.25$ and $0.5$ Mpc, and Table 4 reports the significance of the segregations found. We have only 17 Seyferts versus 76 non-Seyferts, so the statistics are low; nonetheless, the statistical evidence for these segregations is greater than or about $2\sigma$. We have verified that subsets of brighter and brighter galaxies show enhanced small-scale segregation, but the number of Seyferts soon becomes so small that any statistical consequence is doubtful. We note, however, that the imposed limits in $M_B$ are essentially the same as those of the ALSK sample; as a consequence, 15 of the 17 VV91 Seyferts are already included in ALSK, and thus the results just obtained are not independent of the ones found with ALSK.

A number of conclusions can now be drawn : both the CfA and the VV91 samples tell us that no segregation is present for AGNs hosted in $S_{early}$ galaxies less luminous than $M_B \sim -20$; we point out that this is roughly Tully's (1988b) $M^*$ parameter in the luminosity function of NBG galaxies. The enhanced small-scale segregation of Seyferts with respect to LINERs seems convincing, notwithstanding the poor statistics. Moreover, the ALSK sample appears to be a good one for investigating the density segregation of AGNs within the Local Supercluster, as it has the right magnitude limit and its completeness is confirmed by later works.

We have taken a look at the reliability of the Seyfert classification of VV91, by inspecting published spectra in some debated cases; in this way we have selected a sample of 41 *bona fide* Seyferts (within 36 Mpc), the analysis of which, conducted in exactly the same way as VV91, has fully confirmed the above conclusions. Finally, we remember that the VV91 compilation contains also a number of LINER galaxies; adding them to the Seyfert sample, we find the same small-scale behaviour and begin to see some weak large-scale segregation. Again, besides the fact that we have no guarantee of completeness for the LINERs, most of them are already contained in ALSK, so this result is not in fact independent of the one we found before.

## 5 Discussion and Conclusions

The analysis of the environment of Local Supercluster galaxies hosting a low luminosity AGN has allowed us to reach a number of interesting conclusions, which we summarize here: (1) no density segregation is observed for AGNs in ellipticals, lenticulars and late-type spirals, and in early-type spirals less luminous than $M^*$; (2) early-type luminous ($M_B \leq M^*$) spirals hosting a Seyfert or LINER nucleus appear to prefer denser environments on all the scales, from tenths of Mpc to a few Mpc; (3)

LINERs and Seyferts show very similar behaviour, except that Seyferts show enhanced small-scale segregation (4) small-scale segregation is enhanced if we exclude barred galaxies; (5) AGNs in early-type spirals prefer the central zones of the Local Supercluster. The statistical significance of all these segregations is always greater than $2\sigma$ and often $>99\%$; the large-scale segregation is significant at the $3\sigma$ level.

The small-scale segregation found confirms qualitatively the previous results of Dahari (1984), MacKenty (1989) and Rafanelli & Violato (1993); however, our evidence is less drastic than that of Dahari and Rafanelli & Violato, whose analyses are affected by some important biases (discussed in § 1), from which our approach is free because of construction. In this respect the negative result of Fuentes-Williams and Stocke (1986), who paid more attention to the definition of their control sample, is not inconsistent with ours, as weak segregation can be easily destroyed by projection effects. We stress once again that the main observational problem is that in order to detect a significant segregation in an AGN sample we need to include LINERs, which can be easily identified only in bright and nearby galaxies.

The small-scale segregation is fully consistent with the scenario outlined by Osterbrock (1993), according to which Seyfert activity is stimulated by interaction, as discussed in § 1, while LINER activity represents a later phase in which most of the gas has been swallowed by the central SBH, but a low activity still remains. On these bases, we expect both Seyferts and LINERs to stay in denser environments, on scales of tenths of Mpc, with respect to non-AGN galaxies, with Seyferts tending to have nearer companions, as they are the product of the most recent interactions; this is just what we found. It remains to be explained why this mechanism is found to work only in $S_{early}$ luminous galaxies. It is not unexpected that the LINERs in ellipticals and lenticulars have not been stimulated by interactions, as they are hosted in gas-poor systems, but this idea does not explain the absence of small-scale effects on AGNs hosted in gas-rich late spirals; the fact that galaxies less luminous than $M^*$ also show no density segregation suggests that the fundamental parameter that decides the possibility of tidal stimulation of AGN activity is the mass of the galactic bulge, which, for a given morphological type, increases with increasing luminosity and decreasing morphological type.

The fact that a bar inhibits AGN segregation is rather an original finding; we would expect, on the basis of Shlosman et al.'s (1990) considerations, a bar to help the infall of gas into the nucleus, and as a consequence a greater density segregation should be seen in the barred systems. There are at least three possible explanations for this fact: (1) for some unknown dynamical reasons a bar physically inhibits the gas from falling into the nucleus; (2) a bar suffices to make the gas infall toward the nucleus, so that an interaction has no effect and we will see no density segregation in barred systems; (3) as suggested by Hasan & Norman (1990) and by Pfenninger & Norman (1990), the gas inflowed into the nucleus can have a destructive feedback action on the bar, so that interaction-triggered AGNs will be hosted in unbarred spirals.

Superimposed on small-scale segregation we find a strong, $3\sigma$ significant large-scale segregation, which follows the geometry of the Local Supercluster; we stress that it cannot be a simple reflection of the detected small-scale behaviour, as it extends over a distance of at least 20 Mpc (see Fig. 2). It is difficult to understand why a very local phenomenon like nuclear activity is influenced by a large-scale fluctuation.

We suggest a possible explanation: it is known that galaxy morphology depends on the environment, but whether this dependence is local or global is still being debated (see, e.g., the review by Mamon 1993). Santiago & Strauss (1992) have found the whole CfA volume, clusters excluded, to be differently traced by ellipticals and spirals. If we now speculate on the existence of a large-scale bulge-to-disk segregation at fixed morphological type, with more massive bulges staying more probably in large-scale overdensities, our large-scale AGN segregation would be just a reflection of this fact, as AGNs are theoretically supposed to form more easily in deeper gravitational wells (see, e.g., Rees 1993; Haehnelt & Rees 1993). In line with this view, Giuricin et al. (1993b) have found that, for a given morphological type, LINER nuclei have greater central near-infrared luminosity than non-LINERs, which suggests that the former objects have typically more massive bulges than the latters. On the other hand, to our knowledge the only work (Solanes, Salvador-Solé & Sanromà 1989) on the relation of the bulge mass, at fixed morphological type, to the environment is focused on small scales. So, the conjectured dependence of the bulge mass on large-scale density has still to be investigated.

In any case, this large-scale behaviour of AGNs is, as far as we are aware, an original result, and naturally raises the problem of how the AGNs trace the large-scale structure of the universe. For our nearby universe, a study on the large-scale distribution of Seyferts has been carried out by Georgantopoulos & Shanks 1993); by means of a 2-point correlation analysis, they have found a sample of IRAS Seyfert galaxies to be more clustered than IRAS galaxies, on scales of tens of Mpc; however, optical galaxies are more clustered than IRAS ones as well, so that the evidence for enhanced clustering of Seyferts with respect to optical galaxies is not clear. But if Seyferts and LINERs represent different evolutionary phases of the same object, the AGN, and if high-redshift QSOs are just the ancestors of our nearby low-luminosity AGNs, our results could be useful for understanding how to connect the large-scale structure as traced by nearby galaxies with that traced by far-away QSOs.

The authors thank Armando Pisani for his enlightening discussions. The authors are also grateful to Harold G. Corwin, Jr., for having kindly provided them with the ninth tape version of the Third Reference Catalogue of Bright Galaxies. This work was partially supported by the Ministry of University and Scientific and Technological Research (MURST) and by the Italian Research Council (CNR-GNA).

# Tables

**Table 1:** Number of galaxies in the ALSK morphological type subsamples

|         | E+S0 | $S_{early}$ | $S_{late}$ |
|---------|------|-------------|------------|
| AGN     | 22   | 55          | 29         |
| non-AGN | 36   | 29          | 88         |
| total   | 58   | 84          | 117        |

**Table 2:** ALSK $S_{early}$: significance of the differences (in percent) between AGNs and non-AGNs for the $\rho_\sigma$ and $D_V$ distributions ($\sigma$ in Mpc).

|              | N  |    | $\sigma=0.25$ | 0.5    | 1.0   | 2.0   | $D_V$  |
|--------------|----|----|---------------|--------|-------|-------|--------|
| AGN          | 55 | KS | 92.40         | 97.21  | 97.88 | 98.70 | 99.62  |
| vs           | &  | RS | 97.90         | 99.47  | 98.89 | 99.59 | 99.97  |
| non-AGN      | 29 | MW | 97.52         | 99.46  | 98.89 | 99.59 | 99.97  |
| AGN vs       | 47 | KS | 81.32         | 94.85  | 94.10 | 96.38 | 99.30  |
| non AGN      | &  | RS | 93.02         | 97.28  | 94.41 | 98.09 | 99.87  |
| (no Virgo)   | 28 | MW | 91.84         | 97.27  | 94.40 | 98.09 | 99.87  |
| LINER        | 44 | KS | 80.17         | 91.70  | 89.75 | 97.83 | 99.76  |
| vs           | &  | RS | 95.30         | 97.76  | 97.00 | 99.29 | 99.97  |
| non-AGN      | 29 | MW | 94.47         | 97.76  | 97.00 | 99.29 | 99.97  |
| Seyfert      | 11 | KS | 90.74         | 97.20  | 98.75 | 82.74 | 66.91  |
| vs           | &  | RS | 97.05         | 99.54  | 98.11 | 95.60 | 94.75  |
| non-AGN      | 29 | MW | 96.20         | 99.54  | 98.11 | 95.60 | 94.74  |
| AGN          | 55 | KS | >99.99        | 99.99  | 99.99 | 99.69 | 97.14  |
| vs           | &  | RS | 99.99         | >99.99 | 99.98 | 99.92 | 98.37  |
| NBG non-AGN  | 48 | MW | 99.99         | >99.99 | 99.98 | 99.92 | 98.37  |
| AGN vs       | 35 | KS | 99.38         | 99.69  | 99.02 | 99.61 | 99.76  |
| non-AGN      | &  | RS | 99.60         | 99.81  | 99.59 | 99.65 | 99.94  |
| (SA+SAB)     | 19 | MW | 99.50         | 99.81  | 99.59 | 99.65 | 99.94  |

**Table 3:** ALSK $S_{early}$: significance of the differences (in percent) between AGNs and non-AGNs, for the $R_{SG}$, $\Theta_{SG}$ and $Z_{SG}$ distributions.

|         | N  |    | $R_{SG}$ | $\Theta_{SG}$ | $Z_{SG}$ |
|---------|----|----|----------|---------------|----------|
| AGN     | 55 | KS | 95.35    | < 90          | 84.71    |
| vs      | &  | RS | 99.85    | < 90          | 98.04    |
| non-AGN | 29 | MW | 99.85    | < 90          | 98.04    |

**Table 4:** VV91 $S_{early}$: significance of the differences (in percent) between the $\rho_\sigma$ distributions ($\sigma$ in Mpc) of Seyferts and non-Seyferts.

|             | N  |    | $\sigma=0.25$ | 0.5   |
|-------------|----|----|---------------|-------|
| Seyfert     | 17 | KS | 90.52         | 96.14 |
| vs          | &  | RS | 96.29         | 98.72 |
| non-Seyfert | 76 | MW | 95.04         | 98.70 |

# Figure captions

**Figure 1:** $\rho_\sigma$ distributions of ALSK $S_{early}$ galaxies.

**Figure 2:** Virgocentric coordinate distributions of ALSK $S_{early}$ galaxies.

**Figure 3:** $\rho_\sigma$ distributions of VV91 $S_{early}$ galaxies.

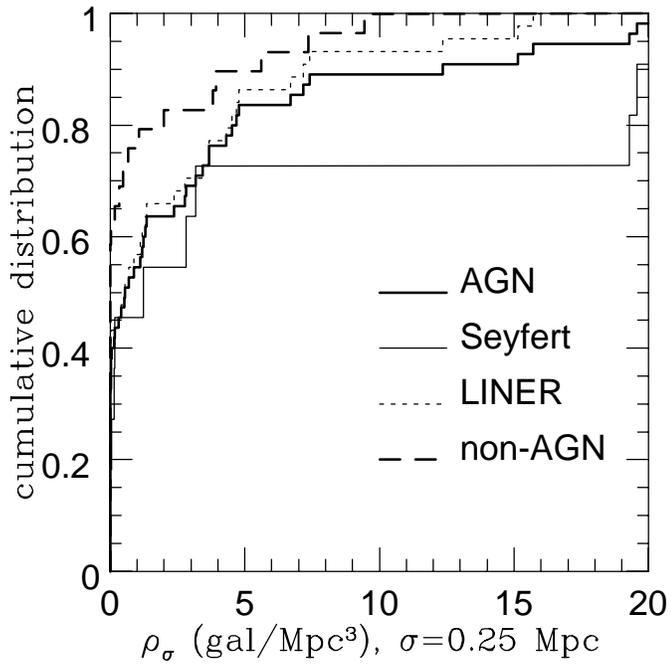
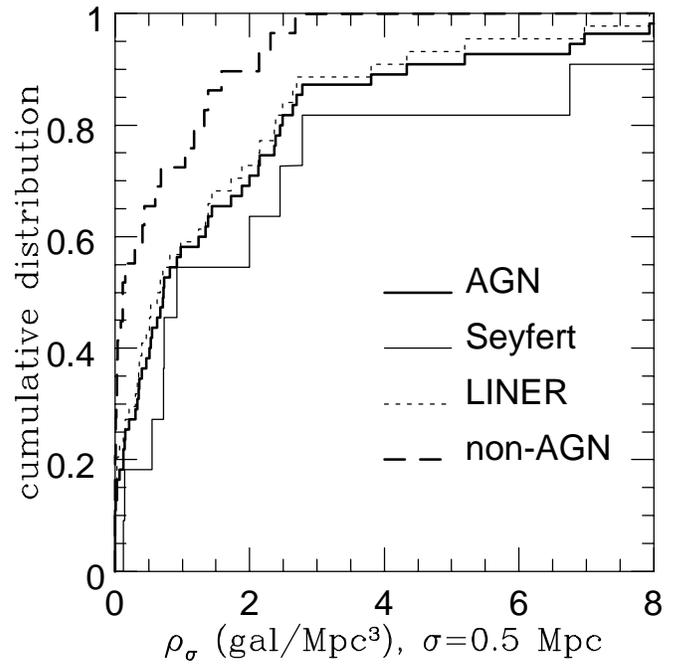
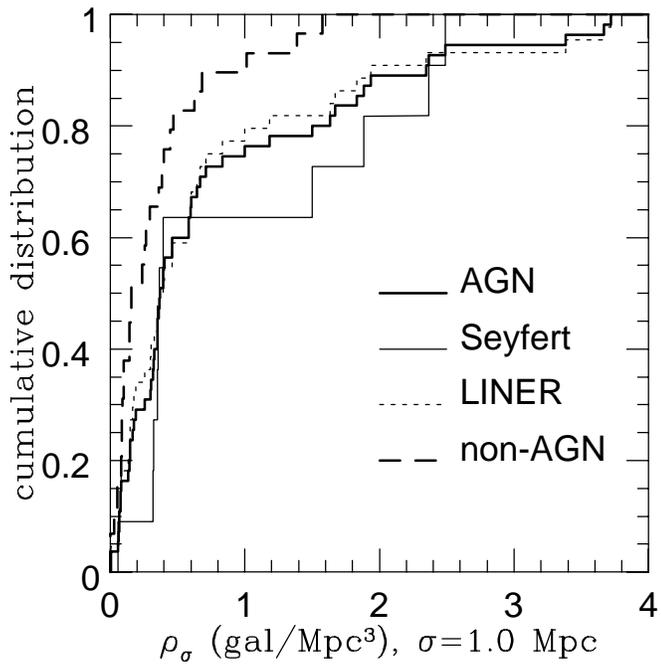
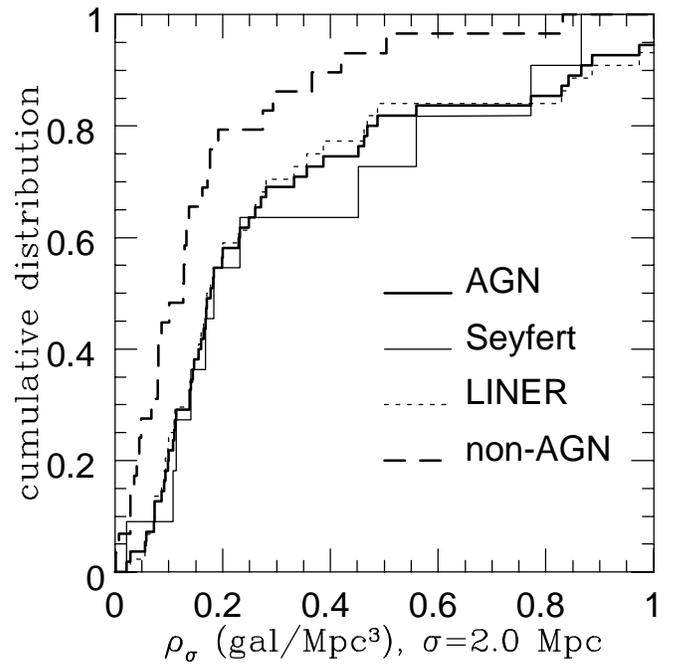

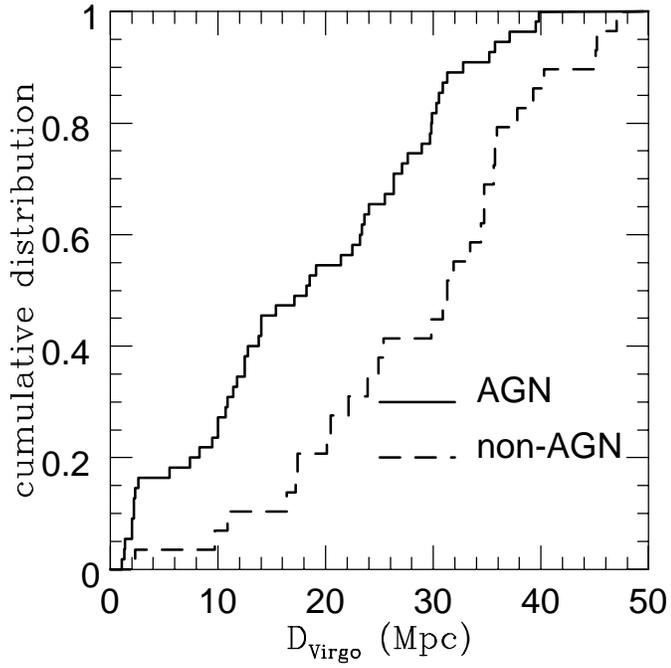
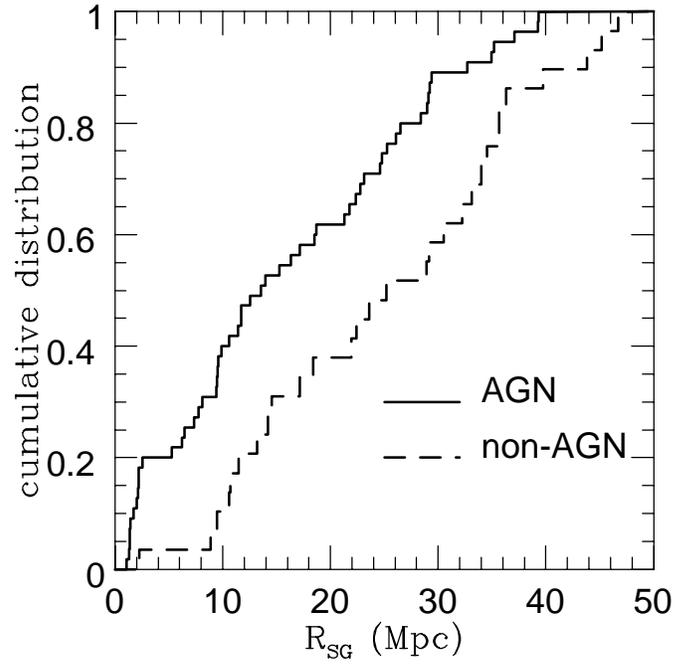
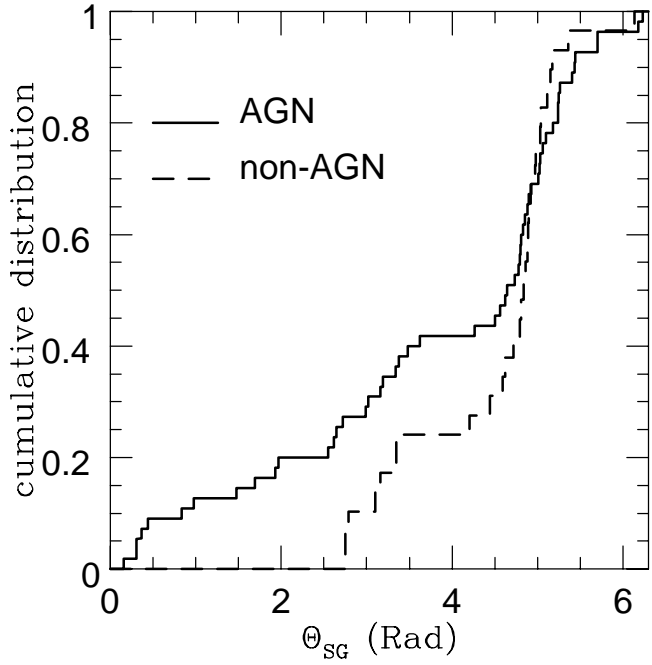
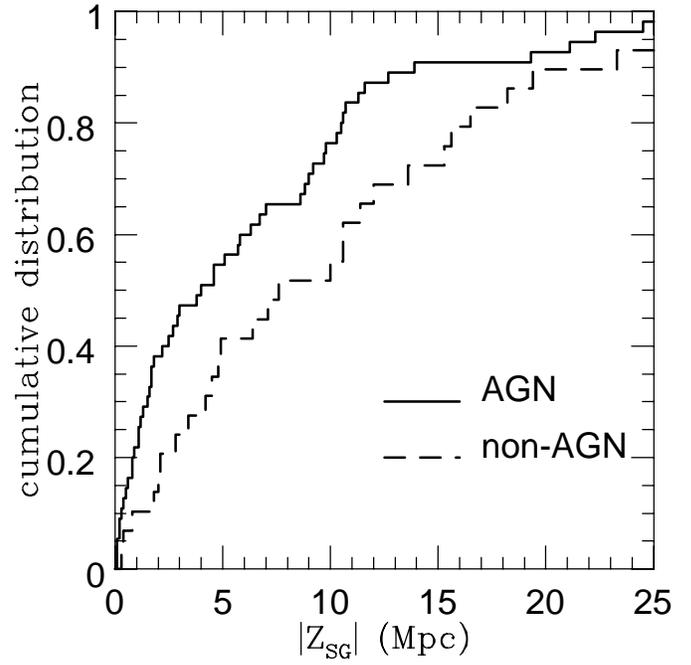

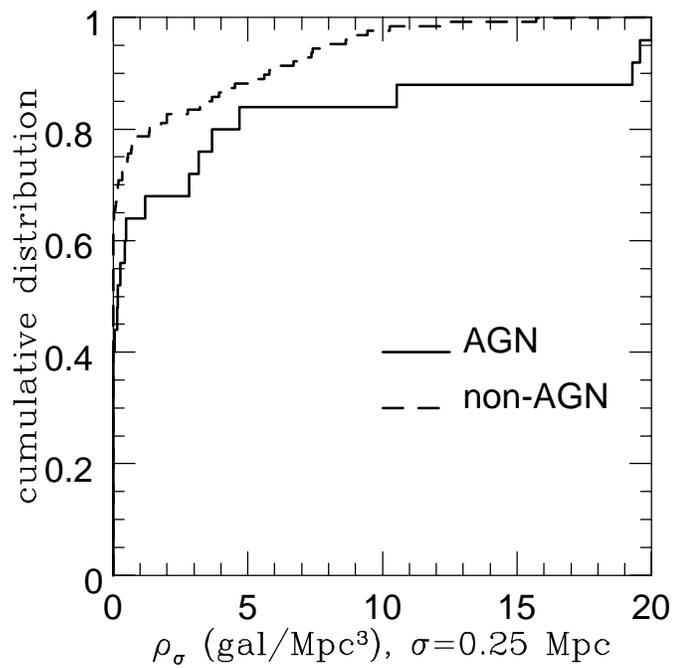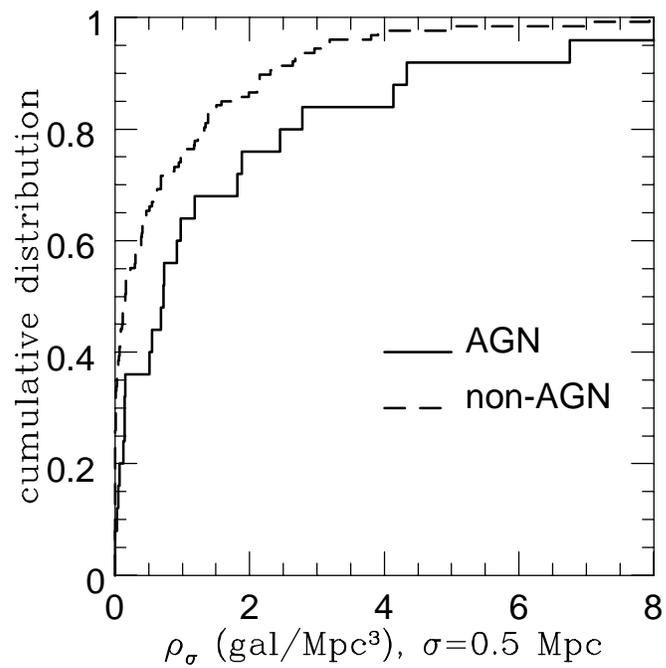